# Environmental permittivity-asymmetric BIC metasurfaces with electrical reconfigurability

*Haiyang Hu[1,4], Wenzheng Lu[1,4], Rodrigo Berté[1], Stefan A Maier[2,3], Andreas Tittl[1*]*

1. Chair in Hybrid Nanosystems, Nanoinstitute Munich, Faculty of Physics, Ludwig-Maximilians-Universität München, Königinstraße 10, 80539 München, Germany.

2. School of Physics and Astronomy, Monash University Clayton Campus, Melbourne, Victoria 3800, Australia.

3. The Blackett Laboratory, Department of Physics, Imperial College London, London SW7 2AZ, United Kingdom.

4. These authors contributed equally

*E-mail: Andreas.Tittl@physik.uni-muenchen.de

Abstract

In the rapidly evolving field of nanophotonics, achieving precise spectral and temporal light manipulation at the nanoscale remains a critical challenge. While photonic bound states in the continuum (BICs) have emerged as a powerful means of controlling light, their common reliance on geometrical symmetry breaking for obtaining tailored resonances makes them highly susceptible to fabrication imperfections and fundamentally limits their maximum resonance quality factor. Here, we introduce the concept of environmental symmetry breaking by embedding identical resonators into a surrounding medium with carefully placed regions of contrasting refractive indexes, activating permittivity-driven *quasi*-BIC resonances (ε-*q*BICs) without any alterations of the underlying resonator geometry and unlocking an additional degree of freedom for light manipulation through actively tuning the surrounding refractive index contrast. We demonstrate this concept by integrating polyaniline (PANI), an electro-optically active polymer, to achieve electrically reconfigurable ε-*q*BICs. This integration not only demonstrates rapid switching speeds, and exceptional durability but also significantly boosts the system's optical response to environmental perturbations. Our strategy significantly expands the capabilities of resonant light manipulation through permittivity modulation, opening avenues for on-chip optical devices, advanced sensing, and beyond.

## 1. Introduction

In modern optical physics, permittivity is one of the principal determinants of light-matter interactions, and constitutes a crucial degree of freedom for designing and engineering optical systems and components.[1–3] The artificial engineering of permittivity at interfaces has led to the development of metasurfaces, revolutionizing nanophotonics by enabling subwavelength light manipulation.[4–6] Among versatile metasurface concepts, bound states in the continuum (BIC) metasurfaces stand out as a powerful platform for realizing high-quality resonance modes,[7,8] demonstrating the transformative potential of permittivity engineering in driving progress in the fields of nonlinear optics,[9,10] sensing,[11,12] and lasing.[13]

BICs have attracted growing attention in nanophotonics due to their strong photon localization at the nanoscale.[7,8] BICs are distinguished into two primary categories, each defined by the distinct mechanism through which the eigenmodes of the structures evade coupling with the radiative continuum. The first category, known as accidental BICs, arises under the Fridrich-Wintgen scenario,[14] where the coupling with radiative waves is suppressed through the tuning of system parameters. The second category, symmetry-protected BICs, results from the preservation of spatial symmetries, such as reflection or rotation, which inherently restrict the coupling between the bound state and the continuum due to symmetry incompatibility.[15,16] True symmetry-protected BICs are theoretical entities of infinitely high-quality factors and vanishing resonance width, which can turn into a *quasi*-BIC ($q$BIC) through the introduction of a finite coupling to the radiation continuum, resulting in a finite quality ($Q$) factor and an observable resonance in far-field spectra.[17] Such coupling can be realized through minor geometric perturbations in the symmetry of resonators within the unit cell, for instance, via changes in the length, height, relative angle, or area of the constituent resonators, etc.[9,12,15,18]

However, the generation of $q$BICs with high $Q$ factor in geometrically modulated systems is limited by the precision of current lithographic techniques, compromising the resonators' performance to a much lower $Q$ factor than the theoretical prediction.[19,20] Moreover, the fixed geometry of the resonators upon fabrication limits the possibilities of harnessing dynamic $q$BICs for potential applications in optical modulation[21,22], dynamic sensor[23], and light guiding.[24,25]

An innovative approach to overcome these challenges involves leveraging the isotropic permittivity of unit cell constituents to induce $q$BICs resonances, a strategy that circumvents the need for precise modifications of geometric asymmetry.[26–29] Here, the asymmetry parameter is governed by the difference in the intrinsic permittivity within the unit cell, allowing for the induction of $q$BIC resonances through perturbations in the permittivity symmetry of resonators.

In this work, by extending this concept, we propose embedding identical resonators into surrounding media with different refractive indexes (RI), effectively introducing an environmental permittivity symmetry breaking to this system. Specifically, we fabricate a metasurface consisting of two identical dielectric nano-rods per unit cell, where one rod is embedded in PMMA with higher RI, contrasting with its counterpart in air. This differential RI surrounding breaks the permittivity symmetry and activates the permittivity-asymmetric $q$BICs (ε-$q$BICs), which can respond actively to further changes in RI asymmetry by altering the environment of the uncovered nanorod. The concept of the environmental ε-$q$BICs also opens the possibility of engineering dynamic BICs for active narrowband applications, such as tunable modulators, sensors, filters, and lasers.[21] As an experimental demonstration, we incorporate polyaniline (PANI), an electrically active conductive polymer known for its large RI variation and fast switching speed,[30–33] for the realization of electrically reconfigurable BICs. This integration, achieved through in-situ polymer growth on the metasurface, ensures mechanical and electrical durability, facilitating rapid and reliable switching of ε-$q$BICs states.[34] Specifically, we successfully engineer the radiative coupling of the ε-$q$BICs by leveraging the electro-optical response of PANI, where the $q$BICs resonance in the transmittance spectra can be switched between the "ON" and "OFF" states with a fast switching speed of 12 ms within low operation voltage range from -0.2 V to +0.6 V. Moreover, we observe a superior cycling stability of over 1000 switching cycles without noticeable degradation. Our reconfigurable ε-$q$BICs metasurfaces platform unlocks a new degree of freedom in manipulating radiation couple of BIC systems, which is unachievable with static geometry symmetry-breaking BIC metasurfaces. This active response to the environmental perturbation exploits the synergy derived from integrating optics and electrolyte fluidics on a single chip, especially in the important visible spectral range, which can be transformative in the development of on-demand flat optics and sensing technologies.[21,35–37]

## 2. Results and discussion

### 2.1 Coupling of bound states to the radiation continuum mediated by environmental permittivity.

For the symmetry-protected BIC metasurfaces, the system of two identical rods in an isotropic environment represents an unperturbed, nonradiative bound state (**Fig. 1a**). Breaking the in-plane symmetry of the unit cell can induce coupling of the bound state to the radiation continuum, resulting in a *quasi*-BIC with a finite $Q$ factor (**Fig. 1b**). Conventionally, this coupling channel is opened through breaking the symmetry of the unit cell geometry, for instance by shortening one of the rods, removing the original $C_2^z$ symmetry and giving rise to the geometry-asymmetric *quasi*-BICs (g-*q*BICs). The asymmetric factor here is proportional to the volume change of one of the resonators (**Fig. 1a**).

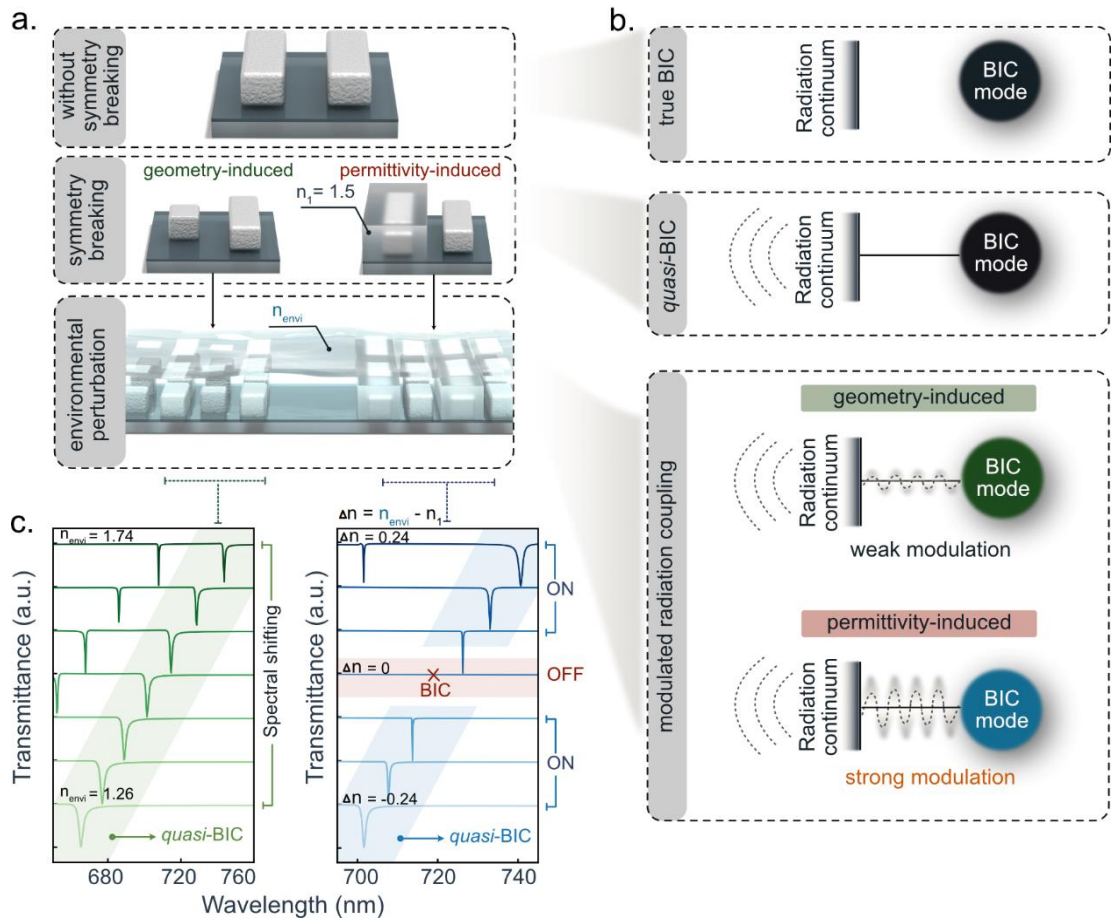

**Figure 1. Strong modulation of radiation coupling in permittivity asymmetric quasi-BIC metasurfaces under environment perturbation. a.** Illustration of the transition from symmetry-protected BIC (top) to *q*BICs (middle) via two distinct approaches: conventional geometry symmetry breaking and innovative permittivity symmetry breaking within the unit cell. Schematic of embedding these two kinds of asymmetric BIC metasurfaces into different environments (bottom), where the refractive index (RI) is variable. The changing of RI can be taken as the environmental

perturbation for the permittivity asymmetric case, leading to different asymmetry parameters. **b.** Schematic illustration of the BIC coupling mechanism. The symmetric geometry with two identical dielectric rods corresponds to a true BIC state, exhibiting no coupling with the radiation continuum (top). Introducing a symmetry-breaking perturbation enables coupling to the far field and transforms the bound state into $q$BICs (middle), which present distinct optical responses to the environmental RI changes depending on the type of symmetry breaking (bottom). **c.** Simulated transmittance spectra reveal that the geometry-induced $q$BICs only present spectral shifts when environmental RI changes, while permittivity-induced $q$BICs respond dynamically and present a transition between the 'ON' and 'OFF' states.

Alternatively, changes in environmental permittivity can disrupt the system's symmetry, thereby enabling the coupling of true BIC modes into the radiative continuum (**Fig. 1a**). In this configuration, two identical nano-rods (within the unit cell) are embedded into heterogeneous surrounding media, each in a different refractive index (RI) environment. Specifically, one nano-rod is placed in a medium with an RI of 1.5 (PMMA), while the other is in the air with an RI of 1.0 (**Fig. 1a**). This RI contrast disrupts their original destructive interference, transitioning the system from a true BIC to an ε-$q$BIC. This change makes it observable in the far field. Here, each meta-atom acts as a Mie-resonant nanoparticle that supports various multipole modes, including electric and magnetic dipoles, as well as higher-order multipole modes[16,38]. The scattering efficiency of each meta-atom is influenced by the characteristics of its surrounding medium. Notably, when these meta-atoms are exposed to different surrounding media, they exhibit distinct scattering behaviors. This scattering effect can be accurately described and thoroughly explained by Mie theory, with the nanosphere as the simplified model.[39,40] Analytical calculating the scattering efficiencies of a pair of nanospheres with different surrounding media (vacuum and PMMA respectively) based on the sphere and core-shell model, reveals that the nanosphere covered by

PMMA exhibits a weaker scattering efficiency compared to the one in vacuum (**Fig. S2**). This difference highlights how the contrasting permittivity of environmental media within the unit cell breaks the system's symmetry.

Geometry-driven $q$BICs (g-$q$BICs) with fixed asymmetry parameters after fabrication, as, e.g., determined by the geometrical differences between two resonators within the unit cell, can naturally also be modulated by surrounding RI changes, this leads only to shifts of the resonance position without direct control over the $Q$ factor (**Fig. 1c**). Therefore, although useful in certain contexts, this mechanism does not fully exploit the potential for dynamic interaction with the surrounding environment. Alternatively, our ε-$q$BICs metasurface design directly and strongly responds to perturbations in the surrounding RI, providing significant advantages over traditional g-$q$BICs metasurfaces. Simulation spectra provide clear evidence that, as the environmental RI transitions from 1.3 to 1.7, the ε-$q$BICs metasurfaces not only exhibit a pronounced shift in the optical response but also demonstrate a unique ability to turn the $q$BIC resonance 'ON' and 'OFF' in the transmission spectrum (**Fig. 1c**). This behavior is directly correlated with the RI contrast between the environments of the different rods, highlighting the superior adaptability of ε-$q$BICs over traditional g-$q$BICs metasurfaces.

## 2.2 Experimental realization of reconfigurable environmental ε-$q$BICs

To construct such ε-$q$BICs metasurfaces, we demonstrate a multi-step nanofabrication approach to obtain a metasurface with the unit cell consisting of two identical rods embedded in distinct surrounding mediums. As shown in **Fig. 2a**, the procedure begins with the spin-coating of PMMA onto pre-fabricated symmetry-protected BIC metasurfaces (see Methods for fabrication details). The crucial step of our fabrication strategy occurs from the second step, where the aim is to construct ε-$q$BICs metasurfaces characterized by alternating rows of $TiO_2$ nano-rods. These rows are distinctively configured, with one set embedded in PMMA and the other exposed to air, thereby introducing a deliberate permittivity asymmetry essential for ε-qBICs functionality. To achieve that, a well-defined lithographic marker system is applied to provide precise spatial alignment.

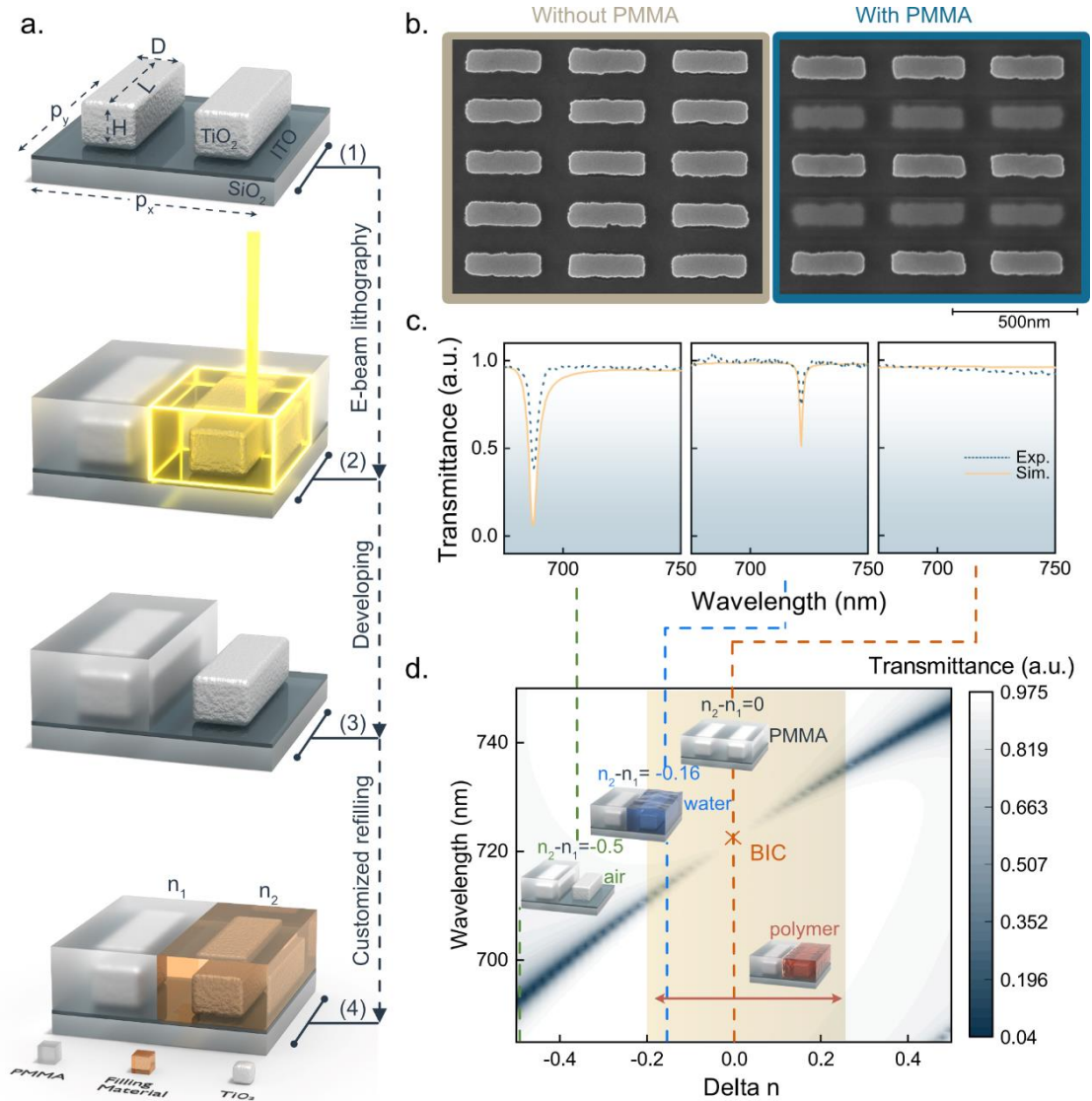

**Figure 2. Reconfigurable permittivity-asymmetry BIC metasurfaces demonstration. a.** Workflow for the fabrication and post-fabrication tuning of the ε-qBIC metasurfaces. The geometrical parameters of the unit cell are: $L$ = 321 nm, $D$ = 107nm, $H$ = 140 nm, $p_x$ = $p_y$ = 403 nm. The thickness of ITO is 50 nm. **b.** SEM images of the fabricated metasurfaces corresponding to the true BICs without permittivity symmetry breaking (before PMMA coating), and the qBICs stemming from a permittivity asymmetry (after PMMA coating). **c.** Experimental and numerical transmittance spectra confirm the high reconfigurability of ε-$q$BICs metasurfaces through customized refilling of different environmental media (air, water or PMMA) for tuning the refractive index contrast ($\Delta n$) surrounding media of the two rods. **d.** Color-coded simulated transmittance map of different ε-$q$BICs metasurfaces as a function of $\Delta n$ and the wavelength. Inserts depict schematic illustrations of unit cells with two rods embedded into different materials. The orange shadowed region showcases the wide tunability of $\Delta n$ provided by the conductive polymer PANI as a surrounding medium (see section 2.3).

Subsequent development of the exposed PMMA regions results in the formation of metasurfaces where each unit cell hosts two identical nano-rods, each within a distinct medium: either PMMA or air. The successful fabrication of PMMA-based ε-$q$BICs metasurfaces is confirmed by scanning electron microscopy (SEM) in **Fig. 2b**. The precise spatial alignment between the two lithographic steps is evidenced by a distinct visual contrast in each unit cell. Specifically, the rod exposed to air displays markedly different features compared to its counterpart encased in PMMA. Here, the PMMA layer is applied with a thickness of 200 nm, sufficient to completely encase the nano-resonators which are 140 nm in height. **Fig. S7** illustrates the influences of PMMA thickness on ε-$q$BICs in transmittance spectra, indicating that saturation gradually occurs after 140nm. Beyond PMMA, this approach is applicable to a variety of resists that offer different RI contrasts (**Fig. S9**), offering on-demand design flexibility.

The asymmetric factor of this PMMA-based ε-$q$BICs metasurfaces, the RI contrast ($\Delta n$), can be actively reconfigured by immersing it in different environment media, such as air ($\Delta n = -0.5$), water ($\Delta n = -0.17$), and PMMA ($\Delta n = 0$), resulting different optical response as depicted in **Fig. 2c**. Specifically, a pronounced ε-$q$BICs was observed in the transmittance spectra with the largest asymmetric factor ($\Delta n = -0.5$). When the asymmetric factor is back to zero (PMMA environment), the quasi-BICs transition to the true BIC state, evidenced by the disappearance of the signal in the transmittance spectra.

In contrast to static alterations of the surrounding medium, the conductive polymer (PANI) enables dynamic tailoring of the optical response of this system (**Fig. 2d**). Its RI can be electrically tailored between 1.3 (oxidized state) and 1.7 (reduced state).[34]

**2.3 Conductive polymer in-situ coating on the ε-$q$BICs metasurface**

As a proof of concept, an electrochemical method was utilized to in-situ grow PANI on the bare resonators, while monitoring the optical transmission response in real-time (**Fig. 3a**). The PANI coating process was initiated through electrochemical polymerization utilizing an aqueous electrolyte. Here, the pre-deposited ITO layer underneath the metasurface functioned as the working electrode, complemented by an Ag/AgCl reference electrode and a Pt wire as the counter electrode, respectively. The polymerization was precisely controlled via a linear scanning voltage from -0.2 V to 0.8 V at a rate of a scanning rate of 25 mV/s. As PMMA covers every second row (odd rows) of the nano-rod resonators, blocking the electrolyte access, resonators in the complementary (even) rows remain exposed to the aniline electrolyte and accessible for a row-selective PANI coating. After 60 cycles of coating, the as-coated PANI achieved a thickness comparable to that of the PMMA (200 nm). The successful integration of PANI onto the metasurface was confirmed through SEM in **Fig. 3b**, which indicates the preferential PANI coating on the even rows of resonators, while the PMMA remains covered on the odd rows.

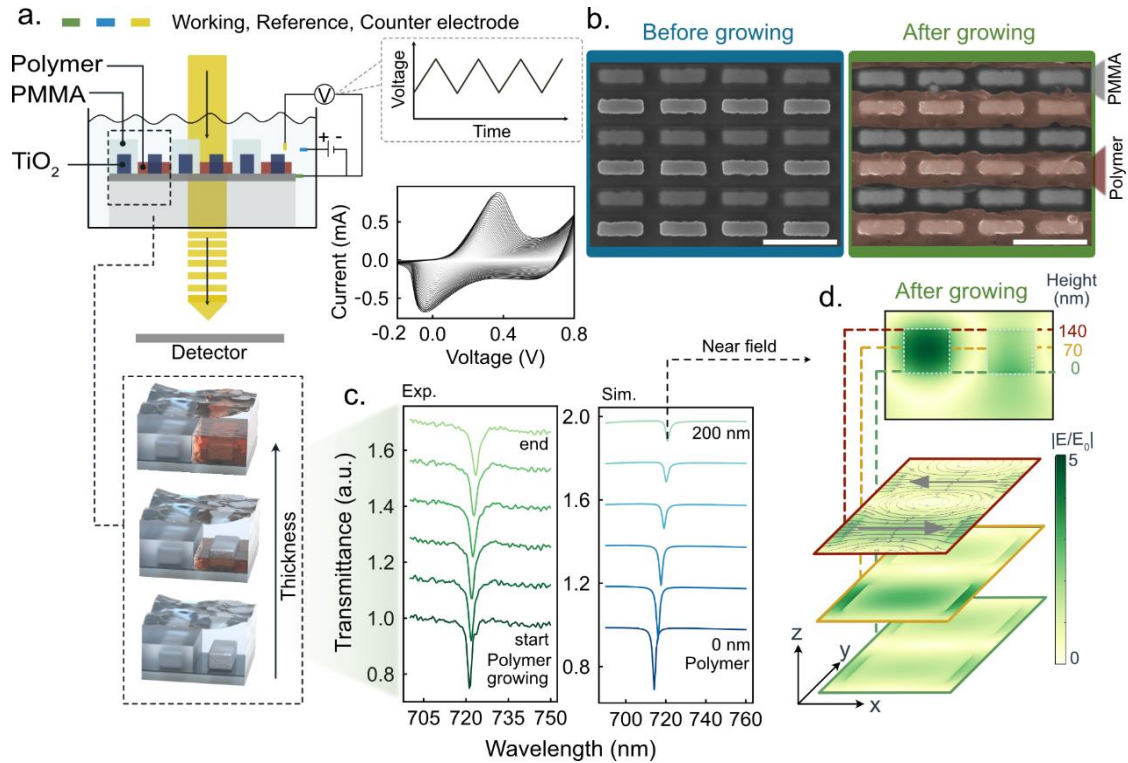

**Figure 3. In-situ coating of conductive polymer on the ε-*q*BICs metasurface. a.** Schematics of the in-situ polymer growing on the ε-*q*BICs metasurfaces in an electrochemical cell. The polymer is grown over a conductive ITO layer without PMMA covering around the rod exposed to the aqueous environment (bottom inset). A scanning voltage from -0.2 to +0.8 V with a total of 60 cycles is applied to polymerize PANI and coat on the nano-rods (right). **b.** SEM images of the ε-*q*BICs metasurfaces before (left) and after (right) in-situ polymer coating (60 cycles). Scale bar: 500 nm. **c.** The experimental in-situ transmittance measurement of ε-*q*BICs metasurfaces with different coating cycles (0, 36, 42, 48, 54, 60) of the PANI at the reduced state (left). The simulated transmittance spectra of ε-*q*BICs metasurfaces with increasing thickness (0, 40, 80, 120, 160, 200 nm) of the PANI (right). **d.** Simulated electric near-fields of individual unit cells with 200-nm thick polymer covering on one of the nano-rods. Top: the side view (*yz*-plane) of the electric field distribution. Bottom: the sliced planer electric field distribution (*xy*-plane) at different heights (0 nm, 70 nm, 140 nm) corresponding to the two-rod resonators unit cell.

This in-situ polymer coating approach allows for the real-time observation of the metasurfaces' optical response under continuous environmental perturbation throughout the PANI coating process. As shown in **Fig. 3c**, increasing PANI thickness on the PMMA-based ε-$q$BICs metasurfaces induces a red shift in the $q$BICs resonance alongside a reduction in modulation depth. This trend agrees well with the numerical simulation, indicating a consistent optical response to the polymer's accretion. Deviations from simulated transmittance spectra primarily stem from the inhomogeneity in the morphology of the coated PANI. After completion of the PANI coating process, the system maintains its $q$BICs state, as evidenced by the $\Delta n$ (asymmetric factor) of -0.2 between PMMA (n = 1.5) with PANI (n = 1.7 in the reduced state). Simulated electric field distribution reveals the asymmetric mode distribution across the two $TiO_2$ nano-rods within each unit cell (**Fig. 3d**). By analyzing $Q$ factors derived from the transmittance spectra of simulations with increasing thickness of PANI, and correlating these with data from experimental in-situ PANI coating cycles, we explore the potential relationship between PANI thickness and its growth cycles (**Fig. S11**).

After integrating PANI into PMMA-based ε-$q$BICs metasurfaces, the system's dynamic reconfigurability can be further explored by electrically adjusting PANI's refractive index.

## 2.4 Electrically reconfigurable ε-$q$BICs based on polymer states switching

Leveraging the advanced electrochemical tunability of PANI, this conductive polymer stands out for its ability to undergo reversible transitions between its reduced and oxidized states (**Fig. 4a**) with large-scale variation in RI (**Fig. S10**).[34] Different RI contrasts between PANI and PMMA ($n_{PMMA}$ = 1.5) can control the radiation loss of this system. The extinction coefficient of PANI increases with the applied voltages,[34] which will also interact with ε-$q$BICs resonance as the intrinsic loss channel.

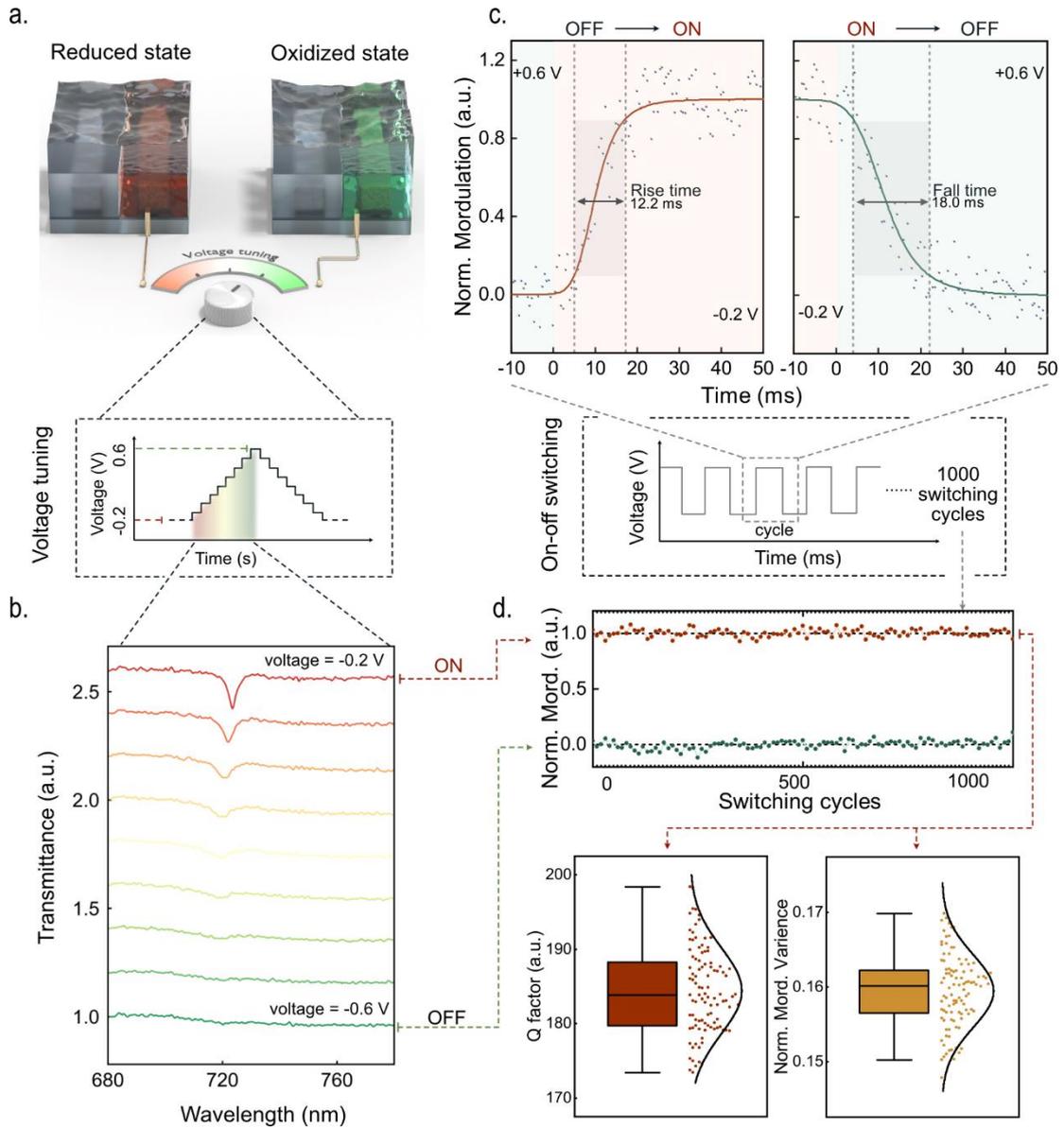

**Figure 4. Electrical tuning optical properties of ε-*q*BICs metasurfaces**. **a.** Schematics illustrating the electrical tuning of the optical response of the ε-*q*BICs. The device is depicted in the reduced and oxidized states, corresponding to an applied voltage of -0.2 V and +0.6 V, respectively. The inset shows the voltage-time profile for tuning the intermediate states of the PANI. **b.** Transmittance spectra at various applied voltages demonstrate the transition of ε-*q*BICs between high (ON) and low (OFF) transmittance modulation. **c.** Time-resolved normalized modulation response of the electrical switching of the ε-*q*BICs. The rise time and fall time, defined by 10% to 90% modulation time for the ON/OFF switching process, are measured to be 12.2 ms and 18 ms, respectively. **d.** Durability analysis of the electrical switching of the ε-*q*BICs.

The upper graph presents the normalized modulation maintenance of the ON/OFF states for over 1000 switching cycles. The lower graphs quantify the variations in the $Q$ factor and modulation of the ε-$q$BICs at the ON state for every 10 switching cycles in the 1000-cycle test, emphasizing its performance consistency.

The ε-$q$BICs resonance modulation strength can be effectively tailored by applying various voltages from -0.2 V to +0.6 V (**Fig. 4b**). Specifically, the ε-$q$BIC resonance modulation strength decreases together with slight blueshifts in wavelength, which is in agreement with our design predictions. We define the ε-$q$BICs with the highest and lowest transmittance modulation as ON and OFF states, corresponding to the applied voltages of -0.2 V, and +0.6 V, respectively. We evaluate the switching speed of our ε-$q$BICs metasurfaces by measuring the time-resolved transmittance response while modulating the applied voltages between the ON (-0.2 V) and OFF (+0.6 V) states, as presented in **Fig. 4c**. The rise time and fall time are defined as the time required for the $q$BIC modulated intensity to rise or fall between the 10% and 90% switching window for switch-on and switch-off processes, respectively, where the modulated intensity is defined as the normalized amplitude of the $q$BIC resonances. The rise time and fall time are measured to be 12.2 ms and 18 ms, respectively, which are mainly limited by the intrinsic material property of the PANI electro-optical dynamics.[34] To demonstrate the durability and repeatability of the electrical switching, we monitor the reconfigurable ε-$q$BICs on our system for 1000 switching cycles. It is clearly observed that the ε-$q$BICs at the ON and OFF states remain fully reversible even after 1000 switching cycles (**Fig. 4d**). The key characteristic supporting excellent durability is that, when electron injection or extraction takes place by voltage application, the highly conductive PANI can redistribute delocalized π-electrons along the polymer chain without structural degradation.[41,42] We further statistically confirm the consistency of the $Q$-factor and modulation of the ε-$q$BICs at the ON states of 100 cycles evenly sampling from the total 1000 switching cycles. Both plots show that the bulk of the data is concentrated around the median, indicating the stability and consistency of the ε-*q*BIC metasurfaces during the electrical switching processes.

## 3. Conclusion

In conclusion, our research introduces a new approach for engineering BIC metasurfaces via permittivity-induced symmetry breaking in the surrounding environment of the resonators, a significant departure from traditional, geometry-reliant methods. This strategy not only overcomes the inherent constraints of geometric asymmetry manipulation but also unlocks the flexibility of post-fabrication control over the coupling to the radiation continuum, a governing aspect of BIC physics. Additionally, our ε-*q*BICs metasurface design showcases enhanced optical response to the refractive index perturbations of the surrounding medium. In contrast to straightforward spectral shifts, our ε-*q*BICs metasurface concept opens possibilities to finely modulate between *quasi* and true BIC states. We have experimentally demonstrated the generation of the ε-*q*BICs from specifically designed metasurfaces, where the coupling channel of BIC to continuum radiation is established by the embedding of two identical $TiO_2$ nano-rods into surrounding media with different permittivity (such as PMMA and air). The ε-*q*BICs concept is further exploited for realizing electrically reconfigurable BICs by integrating a conductive polymer into our metasurfaces. The advantageous electro-optical properties of PANI endow the ε-*q*BICs with excellent electrical reconfigurability. By tuning the optical properties of PANI, the ε-*q*BICs can be dynamically switchable, exhibiting fully reconfigurable *q*BICs with high modulation contrast. Additionally, we have demonstrated the electrical switching performance of the reconfigurable ε-*q*BICs, with fast switching speed (12.2 ms and 18 ms for switch-on and switch-off processes, respectively), and excellent switching durability even after 1000 cycles. Our design is compatible with a range of active materials beyond PANI, including liquid crystals, phase change materials, and transparent conducting oxides. The permittivity-induced BIC metasurfaces unlock an additional degree of freedom for manipulating and engineering BIC states. Thereby laying the groundwork for the on-demand integrated electro-optical devices.

## 4. Materials and Methods

**4.1 Numerical simulations**

Numerical simulations of the permittivity-asymmetric $q$BICs metasurfaces were carried out with CST Microwave Studio with the frequency domain solver. The RIs of PANI in the reduced state and $TiO_2$ were taken from the data of in-house white-light ellipsometry (**Fig. S10**). We utilized the default values implemented in CST Studio Suite for the $SiO_2$ and ITO. The refractive indexes of PMMA were taken from the standard database.[43] The simulations were performed within a rectangular spatial domain containing one unit cell with periodic boundary conditions.

**4.2 Nanofabrication**

We apply sputter deposition (Angstrom) to deposit on the fused $SiO_2$ substrate with multiple layers of materials, which are in sequence 50-nm ITO and 140-nm $TiO_2$. The fabrication of the ε-$q$BICs metasurfaces was based on a three-step electron beam lithography (EBL) process. For the lithography steps, the sample was first spin-coated with a layer of photoresist (PMMA 950K A4) followed by a conducting layer (ESpacer 300Z). We apply eLINE Plus from Raith Nanofabrication with an acceleration voltage of 20 kV and 15 μm aperture to the pattern on the photoresist layer.

In the first patterning process, a 30 nm thick gold marker system was defined on the top of $TiO_2$ film, which was used for aligning the following two fabrication steps. For the second patterning process, based on the gold marker system, we define the position to pattern the metasurfaces with two-rod nanostructures. Subsequent development is carried out in a 3:1 IPA : MIBK (methyl isobutyl ketone) solution for 135s, followed by deposition of a 50 nm chrome layer as a hard mask. Lift-off was conducted in Microposit Remover 1165 overnight at 80 °C. Then, the sample with the hard mask was transferred for reactive ion etching (RIE) of the $TiO_2$ (140 nm) using a

PlasmaPro 100 ICP-RIE from Oxford Instruments. After the RIE etching, the chromium hard mask was removed in a wet Cr etchant. In the last patterning process, we pattern the PMMA layer based on the same marker system on the fabricated BIC metasurfaces with the design of arrays of alternating grooves. Subsequently, we conducted the same developing recipe to clean the patterned region.

**4.3 Electrochemical cell**

The electrochemical polymer coating and electrical switching were carried out in a custom-built electrochemical cell. The electrochemical cell is designed for housing a three-electrode system into a thin layer of aqueous electrolyte with an optical thickness of 1mm on the top of the sample substrate, where a thin transparent glass was used to seal the electrolyte on the top, allowing for an optical transmission measurement through the metasurface. The sample substrate was used as the bottom sealing glass of the cell. The cell also features on the side the in-let and out-let for electrolytes to enable flow-in and flow-out for electrolyte replacement, which is driven by an electrical injector. The sample substrate was connected as the working electrode through a striped metal plate as the electrode contact, whereas a Pt wire and an Ag/AgCl electrode were used as the counter and reference electrodes, respectively. A potentiostat (CHI-760e) is used to apply voltage over time to perform electrochemical polymer coating and switching.

**4.4 In-situ PANI coating and electrical switching**

In-situ PANI coating was carried out by an electrochemical polymerization method, as reported previously.[44] The ITO substrate with the metasurface sample was connected as the working electrode. A cycling voltage in the range from -0.2 V to +0.8 V at a scanning speed of 25 mV/s was applied to the sample in an acidic aqueous electrolyte containing 1 M $H_2SO_4$ and 0.2 M aniline. The thickness of the coated PANI thickness can be controlled by the number of voltage scanning cycles.

For the electrical switching, the electrolyte in the electrochemical cell was replaced by an aniline-free aqueous electrolyte containing only 1 M $H_2SO_4$. To switch the PANI to the oxidized state and reduced state, constant voltages of +0.6 V and -0.2 V, respectively, were applied on the ITO substrate. For electrical switching and cycling, a cycling voltage in the range between -0.2 V and +0.6 V was used.

### 4.5 Optical characterization

The refractive index of PANI was obtained from an ellipsometry measurement on an ellipsometer with dual-rotating compensators and a spectrometer (J.A. Woollam, M2000XI-210). A 100-nm thick PANI file electrochemically prepared on an ITO-coated glass substrate was measured with angle-variable spectroscopic ellipsometry at incident angles of 65º, 70º, and 75º, using a bare ITO-coated glass as blank reference. The transmittance measurements of the metasurfaces were carried out with a Witec optical microscope comprising an air objective (20 X, NA = 0.4, Zeiss, Germany). Illumination was provided by a Thorlabs OSL2 white light source.

## Acknowledgments

This work was funded by the Deutsche Forschungsgemeinschaft (DFG, German Research Foundation) under grant numbers EXC 2089/1 – 390776260 (Germany's Excellence Strategy) and TI 1063/1 (Emmy Noether Program), the Bavarian program Solar Energies Go Hybrid (SolTech), and the Center for NanoScience (CeNS). W.L. acknowledges the Alexander von Humboldt Foundation for the postdoctoral fellowship. S.A. Maier additionally acknowledges the Australian Research Council, and the Lee-Lucas Chair in Physics. Funded by the European Union (ERC, METANEXT 101078018). Views and opinions expressed are however those of the author(s) only and do not necessarily reflect those of the European Union or the European Research Council Executive Agency. Neither the European Union nor the granting authority can be held responsible for them.

## Conflict of Interest

All authors declare no competing financial interest.

## Data Availability Statement

The data that support the findings of this study are available from the corresponding author upon reasonable request.

Supplementary Information for

# Environmental permittivity-asymmetric BIC metasurfaces with electrical reconfigurability

*Haiyang Hu[1,4], Wenzheng Lu[1,4], Rodrigo Berté[1], Stefan A Maier[2,3], Andreas Tittl[1*]*

1. Chair in Hybrid Nanosystems, Nanoinstitute Munich, Faculty of Physics, Ludwig-Maximilians-Universität München, Königinstraße 10, 80539 München, Germany.

2. School of Physics and Astronomy, Monash University Clayton Campus, Melbourne, Victoria 3800, Australia.

3. The Blackett Laboratory, Department of Physics, Imperial College London, London SW7 2AZ, United Kingdom.

4. These authors contributed equally

*E-mail: Andreas.Tittl@physik.uni-muenchen.de

**Tables of contents**

Supplementary Note 1. **Theoretical background for calculating scattering efficiency.**

**Sphere Model for calculating scattering efficiency**

$$Q_{sca} = \frac{2}{x^2} \sum_{n=1}^{n_{max}} (2n+1)(|a_n|^2 + |b_n|^2) \quad (S1)$$

With:

$$a_n = \frac{m^2 j_n(mx)[x j_n(x)]' - \mu_1 j_n(x)[mx j_n(mx)]'}{m^2 j_n(mx)\left[x h_n^{(1)}(x)\right]' - \mu_1 h_n^{(1)}(x)[mx j_n(mx)]'} \quad (S2)$$

$$b_n = \frac{\mu_1 j_n(mx)[x j_n(x)]' - j_n(x)[mx j_n(mx)]'}{\mu_1 j_n(mx)\left[x h_n^{(1)}(x)\right]' - h_n^{(1)}(x)[mx j_n(mx)]'} \quad (S3).$$

Here, $a_n$, and $b_n$ are Mie coefficients for *n*-th order electric and magnetic modes respectively. *m* is the permittivity contrast parameter ($m = \frac{\sqrt{\varepsilon_1 \mu_1}}{\sqrt{\varepsilon_0 \mu_0}}$), where $\varepsilon_0, \mu_0, \varepsilon_1$, and $\mu_1$ are vacuum's, and material's (TiO$_2$) permittivity and permeability. $j_n$ is the spherical Bessel functions of the first kind, and $h_n^{(1)}$ is the spherical Hankel function of the first kind.[1] The analytical calculation is based on the published research.[2]

In the calculation here, we apply the TiO$_2$ as the model material ($n_{core}$ = 2.5) with radius of 200 nm ($R_{core}$).

**Core-shell Model for calculating scattering efficiency**

$$a_n = \frac{j_n[j_n'(m_2 y) - A_n \chi_n'(m_s)] - m_2 j_n'(y)[j_n(m_2 y) - A_n \chi_n(m_2 y)]}{h_n^{(1)}(y)[j_n'(m_2 y) - A_n \chi_n'(m_2 y)] - m_2 h_n^{(1)'}(y)[j_n(m_2 y) - A_n \chi_n(m_2 y)]} \quad (S4)$$

$$b_n = \frac{m_2 j_n(y)[j_n'(m_2 y) - B_n \chi_n'(m_2 y)] - j_n'(y)[j_n(m_2 y) - B_n \chi_n(m_2 y)]}{m_2 h_n^{(1)}(y)[j_n'(m_2 y) - B_n \xi_n'(m_2 y)] - h_n^{(1)'}[j_n(m_2 y) - A_n \chi_n(m_2 y)]} \quad (S5)$$

With:

$$A_n = \frac{m_2 j_n(m_2 x) j_n'(m_1 x) - m1 j_n'(m_2 x) j_n(m_1 x)}{m_2 h_n^{(1)}(m_2 x) j_n'(m_1 x) - m_1 h_n^{(1)'}(m_2 x) j_n(m_1 x)} \quad (S6)$$

$$B_n = \frac{m_2 j_n(m_1 x) j'_n(m_2 x) - m_1 j_n(m_2 x) j'_n(m_1 x)}{m_2 \chi'_n(m_2 x) j_n(m_1 x) - m_1 j'_n(m_1 x) \chi_n(m_2 x)} \quad (S7)$$

and:

$$x = \frac{2\pi R_{core}}{\lambda}, y = \frac{2\pi R_{shell}}{\lambda}, m_1 = \frac{n_{core}}{n_{medium}}, m_2 = \frac{n_{shell}}{n_{medium}}$$

$$\chi_n(x) = -x\sqrt{\frac{\pi}{2x}} N_{n+1/2}(x)$$

In the calculation here, we apply the TiO$_2$ as the core material ($n_{core}$ = 2.5) with radius of 200 nm ($R_{core}$), and PMMA as the shell material ($n_{shell}$ = 1.5) with the thickness of 80 nm ($R_{shell}$). $N_{n+1/2}$ is the Neumann functions which are solutions to the spherical Bessel's equation and represent the spherical waves that diverge from the origin. The analytical calculation is based on the published research.[2]

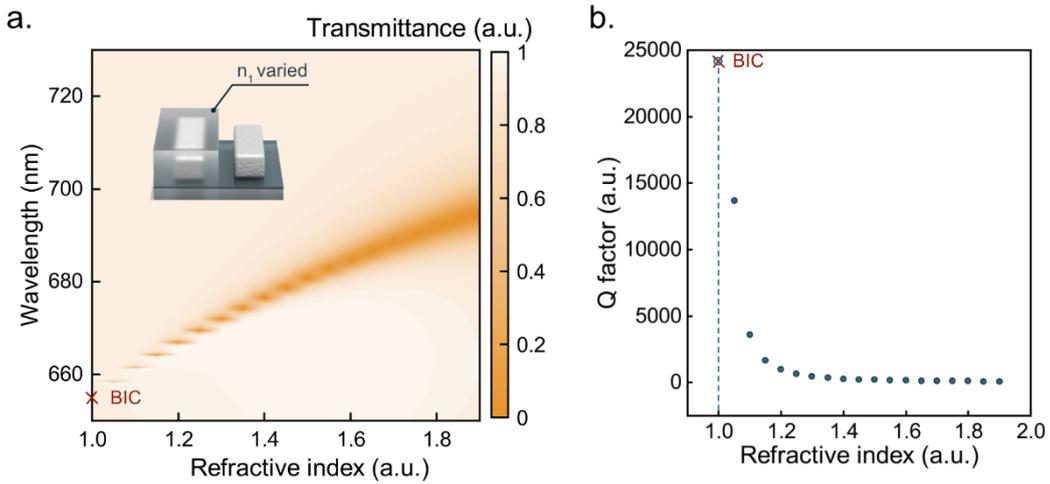

Figure S1. **Numerical simulation of the ε-qBICs metasurfaces with varied $n_1$. a.** Color-coded simulated transmittance map of the ε-qBICs metasurfaces as a function of the wavelength and the refractive index of the surrounding medium on one row ($n_1$). Insert schematic shows the unit cell of the ε-qBICs metasurface, where $n_1$ is variable while $n_2$ = 1. **b.** The quality factor (Q factor), of the ε-qBIC resonance extracted from the transmittance spectra.

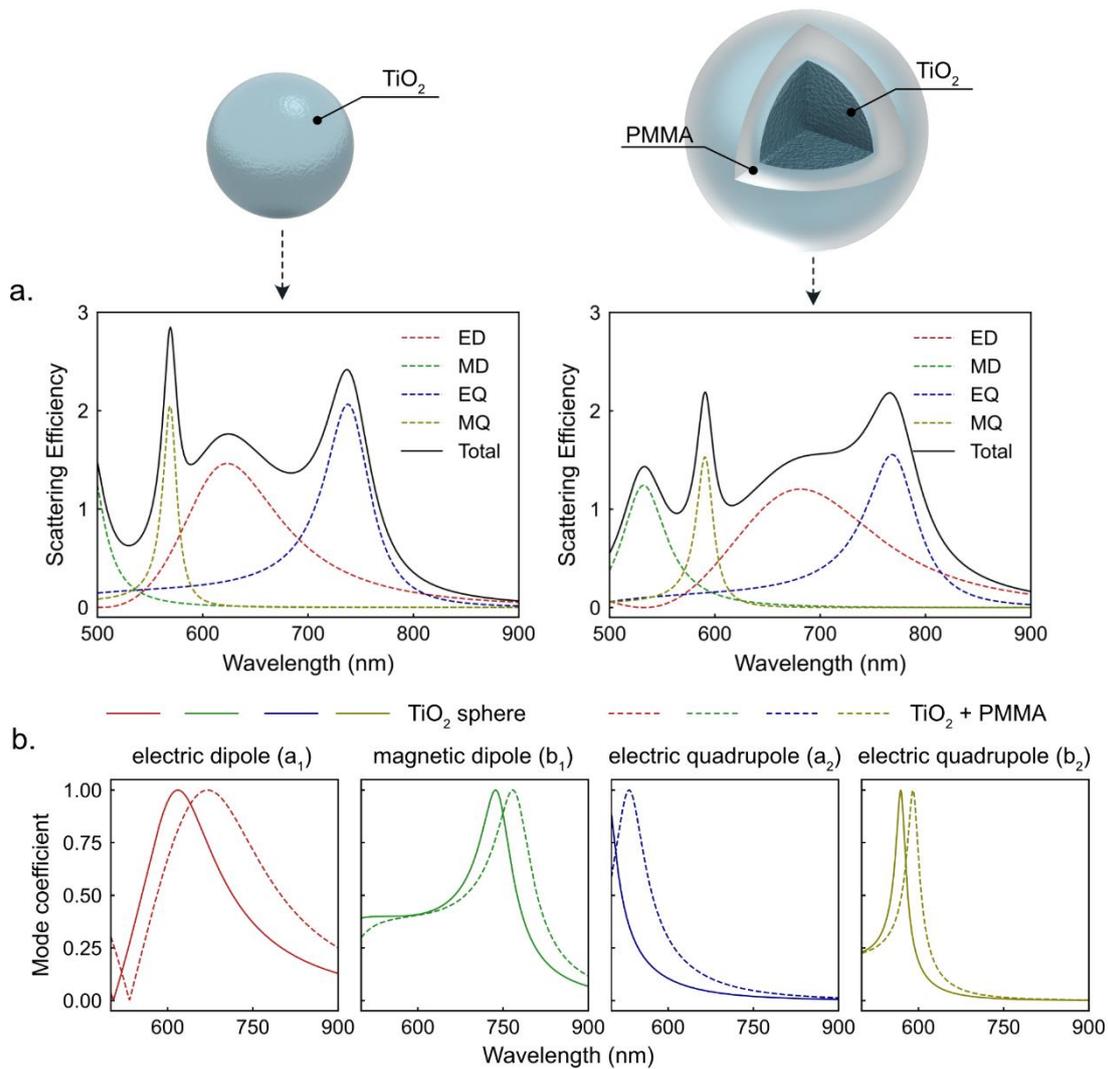

Figure S2. **Analytical calculation of scattering efficiencies. a.** Scattering efficiency across wavelengths ranging from 500 to 900 nm for a standalone $TiO_2$ sphere with a radius of 200 nm (left) based on the Mie sphere model, and a $TiO_2$ sphere (same size) embedded in a PMMA (80 nm) with the core-shell model (right). The contributions from the electric dipole (ED), magnetic dipole (MD), electric quadrupole (EQ), and magnetic quadrupole (MQ) are shown (dash lines), along with the total scattering efficiency. **b.** Individual mode coefficients, including the electric dipole $a_1$, magnetic dipole $b_1$, electric quadrupole $a_2$, and electric quadrupole $b_2$, corresponding to the respective resonances observed in the scattering efficiency spectra, which highlights the influence of the surrounding medium on the optical resonances of the $TiO_2$ spheres.

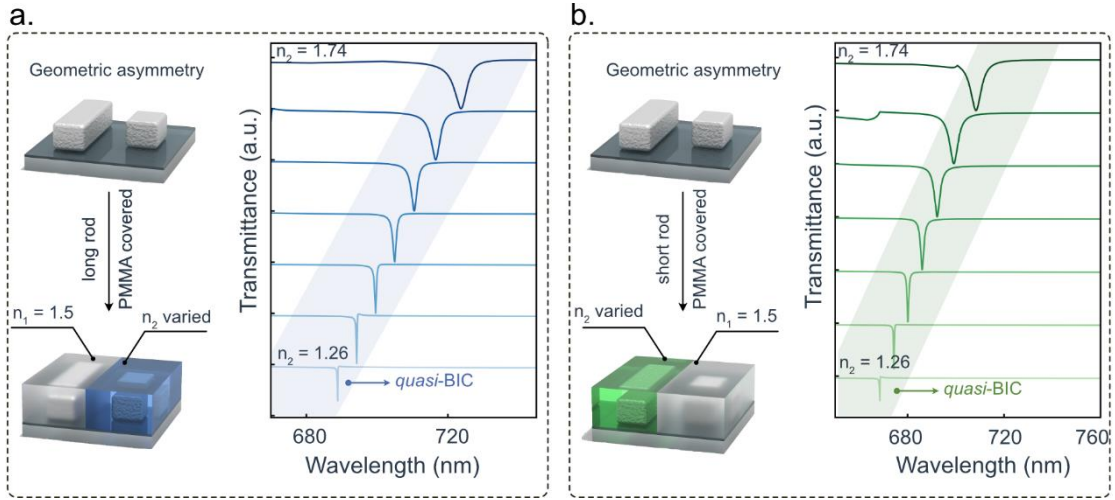

Figure S3. **Numerical simulation of *q*BIC resonances with varied refractive index of surrounding medium on geometry-asymmetric unit cells. a.** Schematics of the unit cell by covering the long rod with PMMA, with the shorter rod exposed to a medium of varying refractive index (left). Alongside are the corresponding to the simulated transmittance spectra (right). **b.** Schematics of the unit cell by covering the short rod with PMMA (left), and their corresponding simulated transmittance spectra (right).

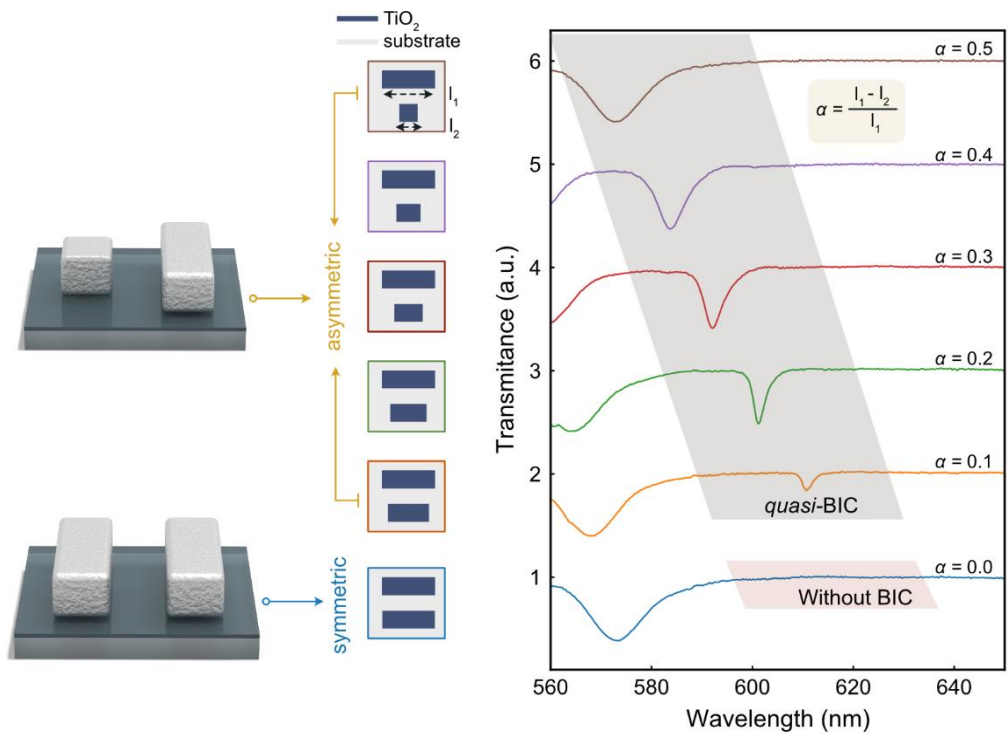

Figure S4. **Geometry induced symmetry breaking of *q*BIC metasurfaces.** The geometry asymmetry in the TiO$_2$ two-rod unit cell is realized by shortening the length

of one of the rod, as shown in the schematics (left). The geometric asymmetry $\alpha$, is defined by $\alpha = \frac{l_1 - l_2}{l_1}$, where $l_1$ and $l_2$ are the length of the longer rod and shorter rod, respectively. The experimentally measured transmittance spectra of the $q$BIC metasurfaces with different $\alpha$ are shown (right). When the unit cell is geometrically symmetry, the BIC is symmetry-protected and no BIC resonance is displayed in the spectrum. The $q$BIC emerges with the presence of geometric asymmetry. When the geometric asymmetry $\alpha$ is increased, the $q$BIC blueshifts and broadens in the peak width.

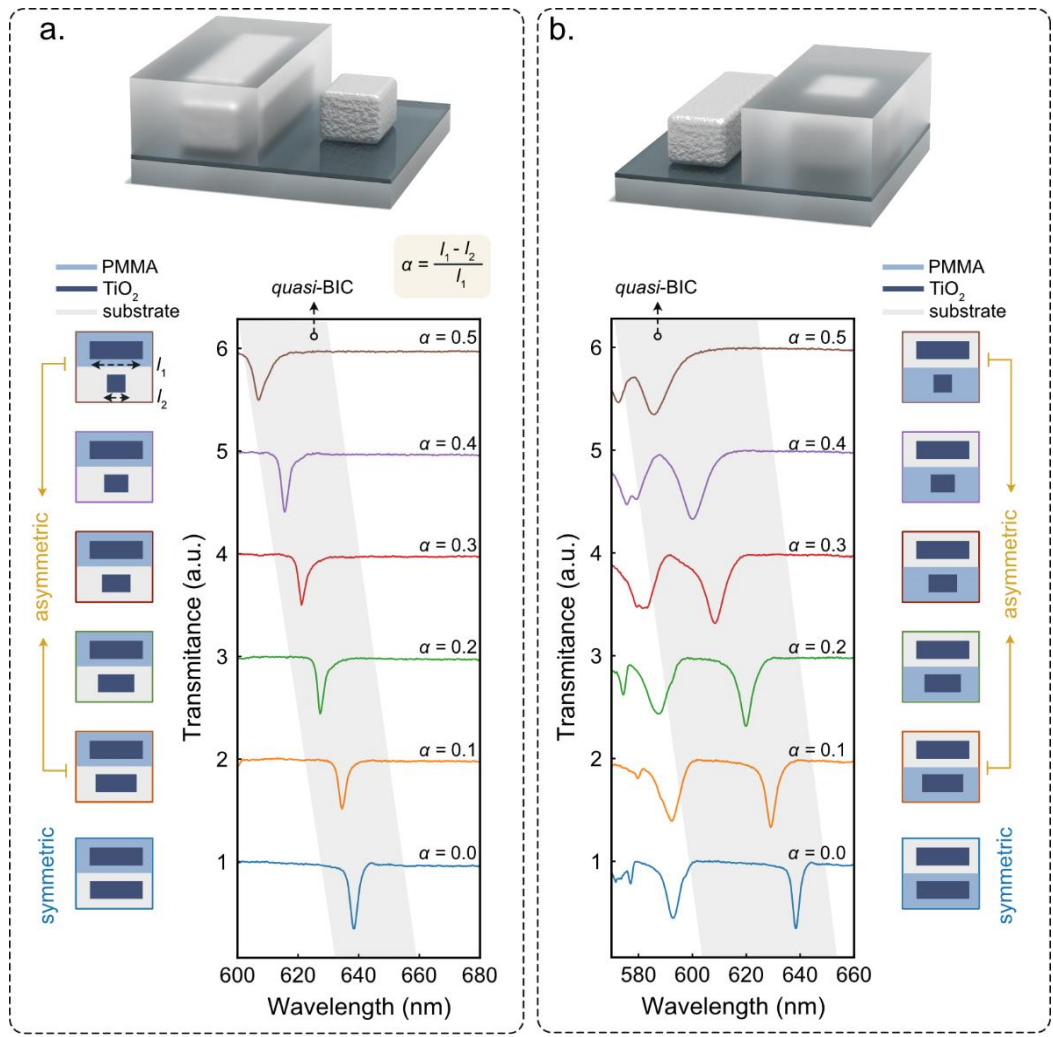

Figure S5. ***q*BIC resonances based on metasurfaces with geometric asymmetry and permittivity asymmetry. a.** $q$BIC metasurfaces with PMMA covered on the long rod, while the short rod is exposed to air. The geometric asymmetry $\alpha$, is defined by the ratio of the length difference of the two rods to the length of the long rod, as shown by the equation in the figure. **b.** *q*BIC metasurfaces with PMMA covered on the short rod, while the long rod is exposed to air. For both cases, as the geometric asymmetry increases, the *q*BIC blueshifts.

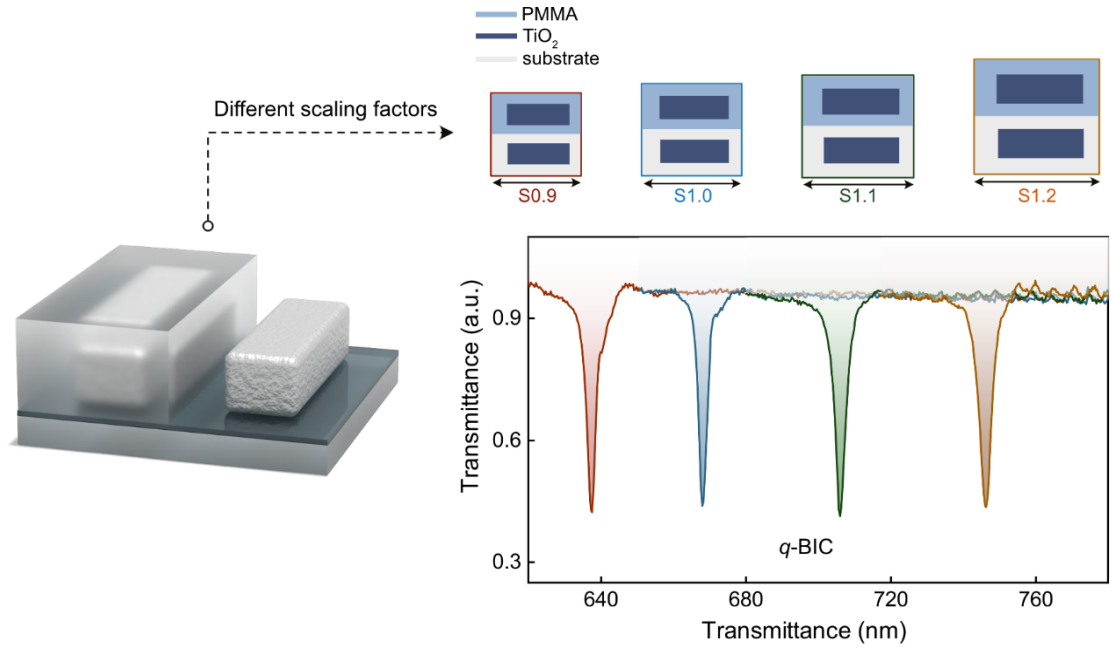

Figure S6. **ε-*q*BIC metasurfaces with varied scaling factor.** Schematics and transmittance spectra of the ε-*q*BIC metasurfaces with varied scaling factors from 0.9 to 1.2. As the scaling factor is increased, the corresponding generated ε-*q*BIC redshifts.

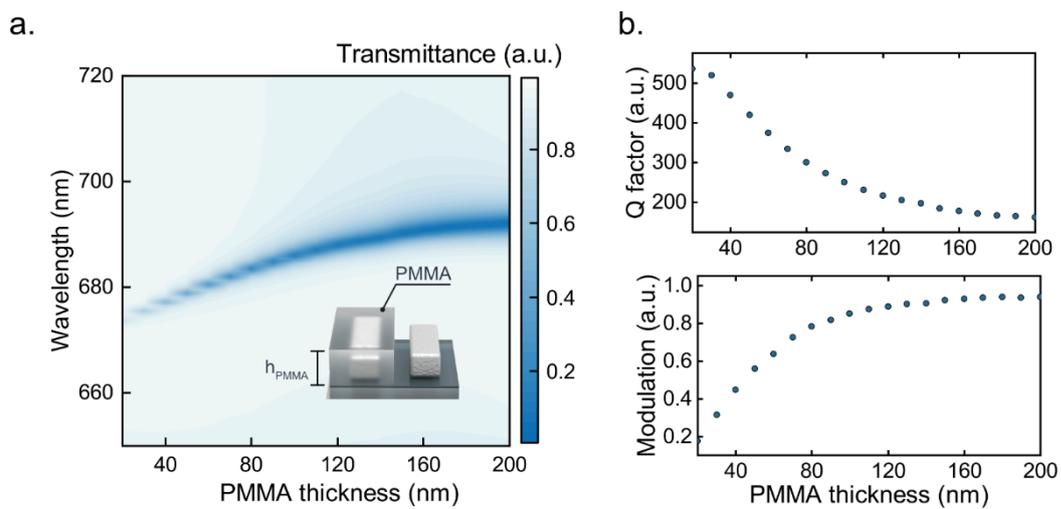

Figure S7. **Numerical simulation of ε-*q*BIC metasurfaces with varied $h_{PMMA}$. a.** Color-coded maps for transmittance spectra of ε-*q*BICs metasurfaces as a function of wavelength and PMMA thickness ($h_{PMMA}$). Inserted schematic shows the unit cell of the ε-$q$BIC metasurfaces, where $h_{PMMA}$ is varied. The PMMA covers on one rod in the unit cell, while the other rod is exposed to air. **b.** Analysis of the $Q$ factor (top) and the modulation (bottom) of the $q$BICs resonances, derived from transmittance spectra across PMMA coating thicknesses, presented as upward and downward trends.

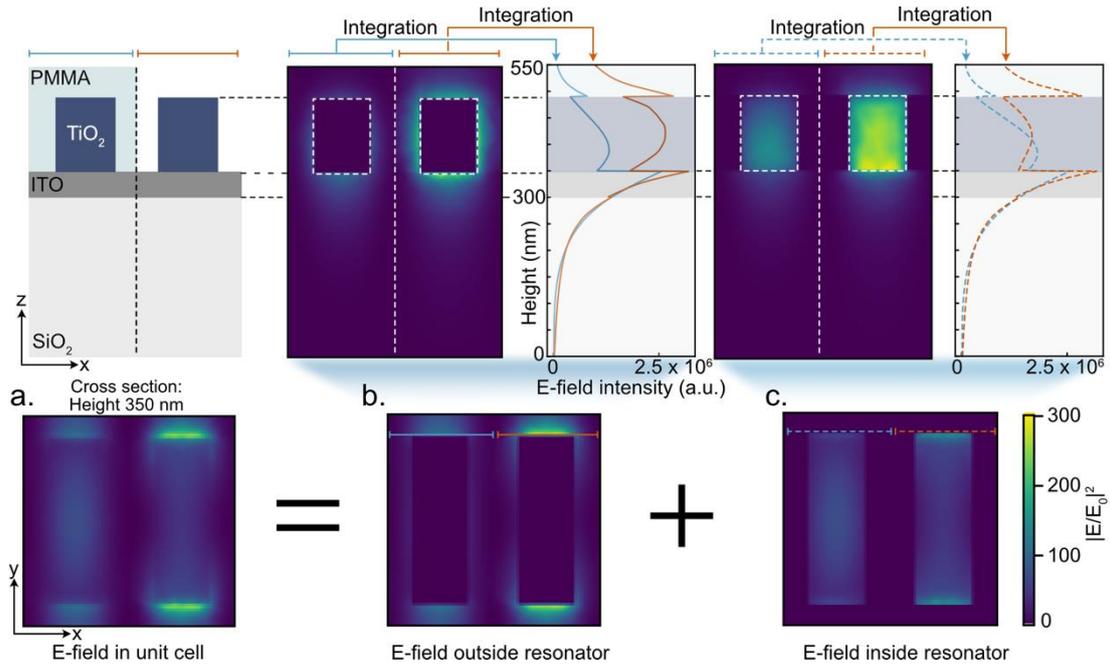

Figure S8. **Simulated electric field distribution of the unit cell and the regional integration analysis. a.** The E-field distribution of the unit cell at the bottom of the $TiO_2$ resonators, where the left $TiO_2$ rod resonator is embedded with PMMA and the right $TiO_2$ rod resonator is exposed to air. The E-field in the unit cell can be divided into two parts, including **b.** the E-field outside the $TiO_2$ two-rod resonators, and **c.** the E-field inside the $TiO_2$ two-rod resonators. The top figures show the integrated E-field distribution profiles along the $z$-axis direction. The e-field intensities of both outside and inside of the uncovered $TiO_2$ rod are generally higher than the one covered by PMMA. The asymmetry of the e-field distribution gives rise to radiation coupling to the far field, exhibiting strong ε-$q$BIC resonance.s

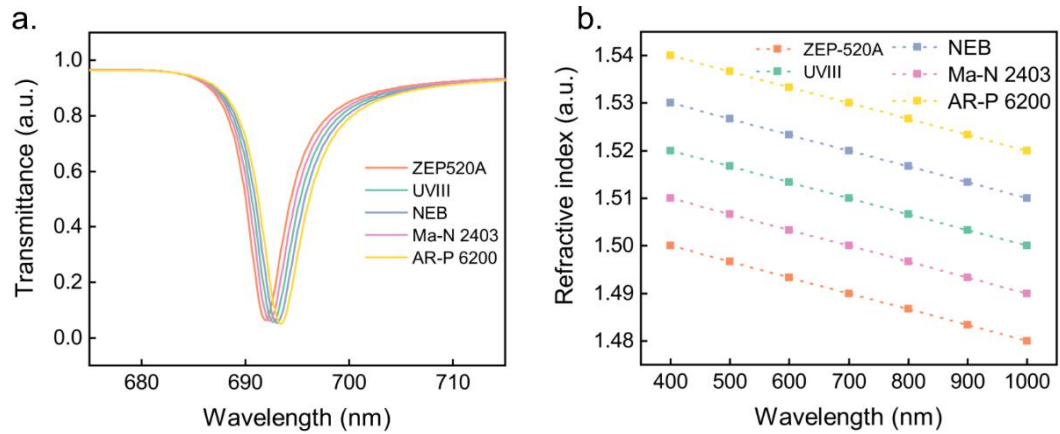

Figure S9. **Simulated transmittance spectra of ε-*q*BICs metasurfaces using different photoresists. a.** Simulated transmittance spectra of the ε-*q*BICs metasurfaces, where one rod in the unit cell is covered by different photoresists with the thickness of 200 nm. **b.** The refractive indices of the e-beam photoresists with linear interpolation used in the simulation.

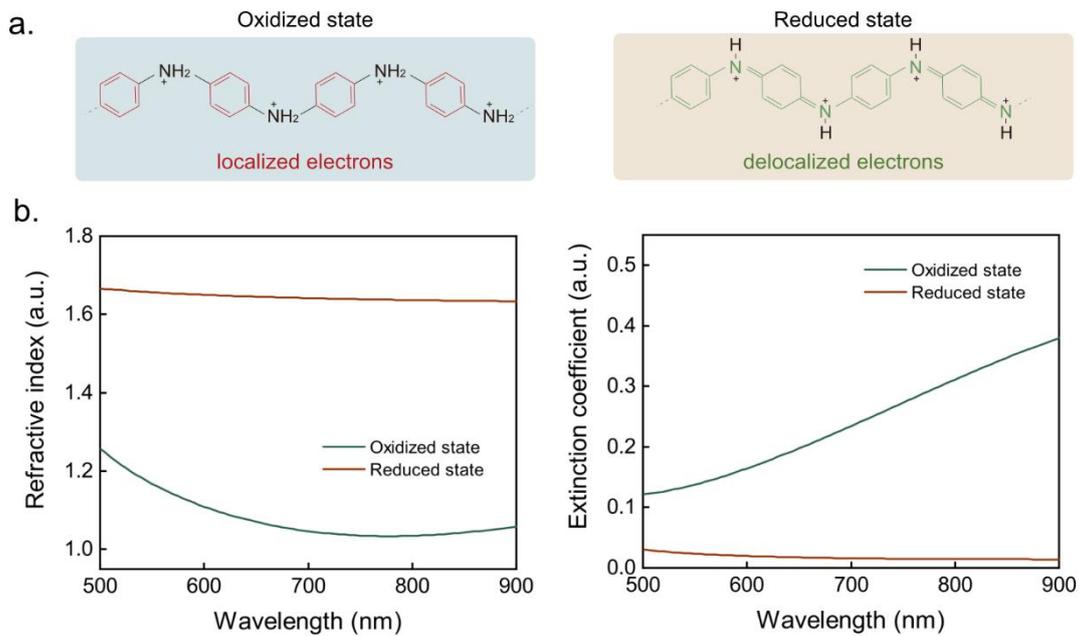

Figure S10. **Measured refractive index and extinction coefficient of PANI in different states. a.** Molecular structure of PANI changes in different states, where the π-electrons are localized in the oxidized state, and delocalized in the reduced state. **b.** The refractive index (left) and extinction coefficient (right) of PANI in oxidized and reduced states.

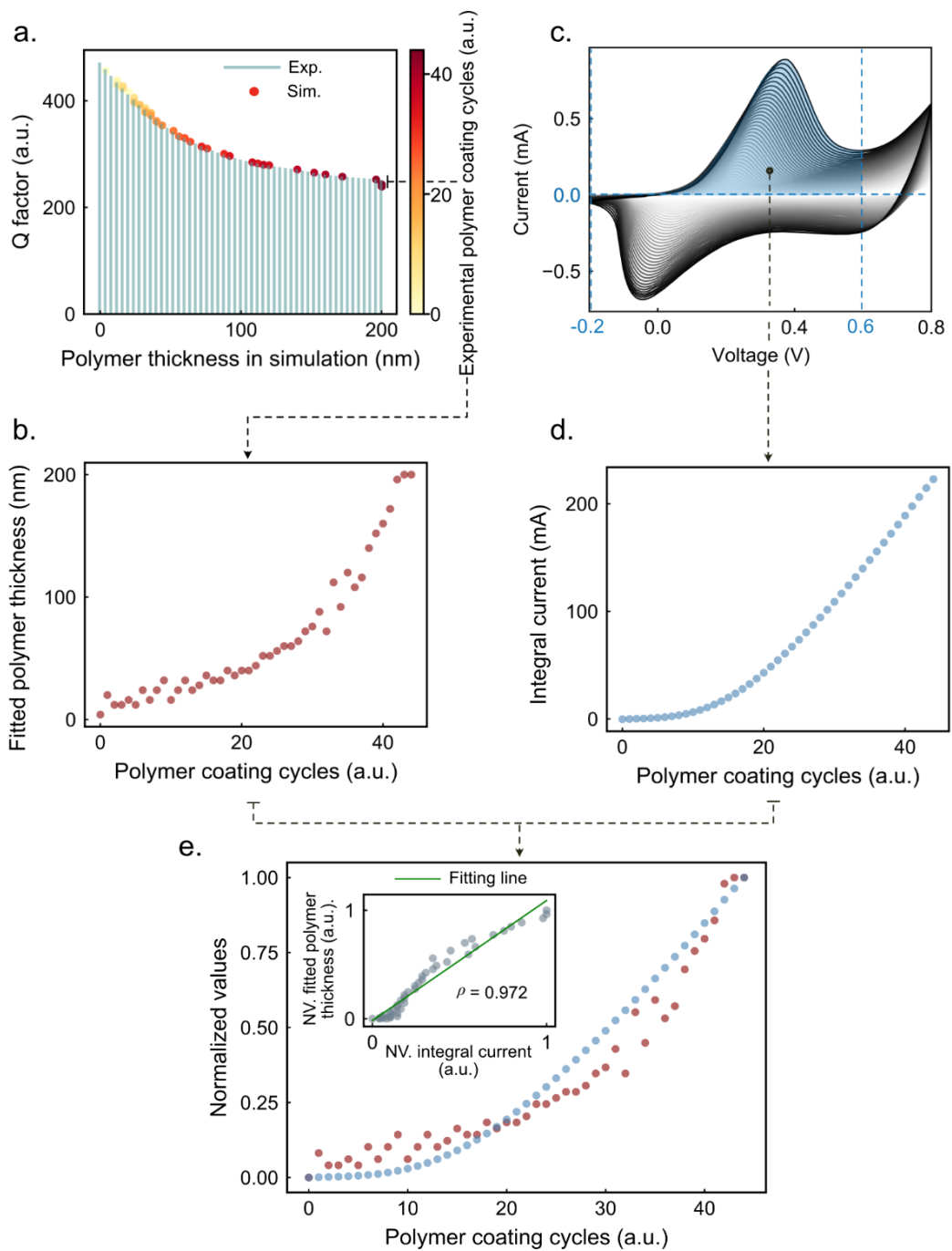

Figure S11. **Analysis of PANI thickness during the in-situ coating process. a.** $Q$ factor analysis for the PANI thickness in the simulation and the PANI coating cycle in the experiment. The $Q$ factors obtained from the simulated and experimental results show excellent agreement. Thus, the $Q$ factors can be used to evaluate the PANI thickness during the in-situ coating process. **b.** The fitted PANI thickness for different coating cycles. The polymer thickness fitting is based on the $Q$ factors obtained in the simulation. **c.** Electrochemical current-voltage diagram recorded during the in-situ

PANI coating process. **d.** Electrochemical integral current for different coating cycles during the in-situ coating process. The integral current is obtained in the range of applied voltage from -0.2 V to +0.6 V in the oxidation process for each polymer coating cycle, reflecting the total amount of PANI coated on the metasurfaces. **e.** Comparison of the fitted PANI thickness and the integral current at different coating cycles. Inserted figure shows the analysis of the discrepancy between the $Q$-factor-based fitted PANI thickness and the integral current. The fitted polymer thickness obtained from the $q$BIC resonance $Q$ factors shows a similar growing trend to the total amount of coated polymer on the substrate. The polymer thickness analyses from both optical BIC resonance and electrical measurement are in good agreement, strengthening the reliability of the PANI thickness evaluation during the in-situ coating process. Therefore, through monitoring the $Q$ factor of the ε-$q$BICs, it further provides insights of the variation of the surrounding media of the resonators.